\documentclass{elsart}
\usepackage{epsfig}

\def\norc #1 {\ {\Vert} #1 {\Vert_{c}}}
\def\norN #1 {\ {\Vert} #1 {\Vert_{N}}}
\def\nor #1 {\ {\Vert} #1 {\Vert}}

\def \Ninf#1 {  {\Vert {#1} \Vert_\infty } }
\def \NLip#1 {  {\Vert {#1} \Vert_\theta } }
\def \Norm#1 {  {\Vert {#1} \Vert } }

\usepackage{amssymb}

\begin{document}

\begin{frontmatter}



\title{Spectral Characterization of Anomalous Diffusion of a 
Periodic Piecewise Linear Intermittent Map}


\author{S. Tasaki}
\address{Advanced Institute for Complex Systems
and 
Department of Applied Physics,\\
School
of Science and Engineering,
Waseda University,\\
3-4-1 Okubo, Shinjuku-ku,
Tokyo 169-8555,
Japan}

\author{P. Gaspard}
\address{Center of Nonlinear Phenomena and Complex Systems, \\
Universit\'e Libre de Bruxelles, Campus Plaine
C. P. 231, \\ B-1050 Brussels, Belgium}

\begin{abstract}
For a piecewise linear version of the periodic 
map with anomalous diffusion,
the evolution of statistical averages of a class of 
observables
with respect to piecewise constant initial densities
is investigated and generalized eigenfunctions of 
the Frobenius-Perron
operator are explicitly derived.
The evolution of the averages is 
controlled by real eigenvalues as well as continuous 
spectra terminating at the unit circle. 
Appropriate scaling limits are shown to give
a normal diffusion if the reduced map is in the 
stationary regime with normal fluctuations, a L\'evy
flight if the reduced map is in the 
stationary regime with L\'evy-type fluctuations
and a transport of ballistic type if the reduced 
map is in the non-stationary regime. 
\end{abstract}

\begin{keyword}
generalized spectral decomposition \sep
generalized eigenfunctions \sep
intermittent map \sep anomalous diffusion \sep
L\'evy flight
\PACS 
\end{keyword}
\end{frontmatter}


\newpage


\section{INTRODUCTION}

The Frobenius-Perron (FP) operator is a powerful tool to investigate 
statistical properties of dynamical systems\cite{RuelleTher,PGtext,Dorftext}.
For strongly chaotic maps where all trajectories are 
exponentially unstable such as expanding and hyperbolic systems, 
correlation decays exponentially and the decay rates, 
known as Pollicott-Ruelle resonances\cite{Pollicott,Ruelle},
are given as the logarithms of the zeros of the dynamical
zeta function.  Note that the zeta function is essentially
the Fredholm determinant of the FP operator and its zeros are 
eigenvalues of the FP operator in a generalized sense.
Similar characterization was observed for periodic strongly 
chaotic maps with deterministic 
diffusion (see \cite{PGtext,Dorftext,DetDif2,DetDif3} and references 
therein): 
There, because of 
the periodicity of the system, the FP operator is decomposed into 
Fourier components and each component admits wavenumber-dependent
eigenvalues including hydrodynamic modes of diffusion. 
Moreover, the corresponding eigenfunctions have fractality,
which is closely related to the underlying phase space structure
and the transport coefficients\cite{GDG}.

The method of the dynamical zeta function is extended to certain 
non-hyperbolic maps with marginal fixed points\cite{CasatiArtuso,Artuso}. 
For those maps, the power-law decay of correlation and/or the 
anormalous growth of the mean square displacement are obtained
from the zeta function. The main difference in the zeta function 
between the non-hyperbolic and hyperbolic maps is the existence 
of a cut terminating at the unit circle in the former case. 
In other words, the FP operator of the non-hyperbolic map
has a continuous spectrum terminating at the unit circle. 
However, its spectral structures such as the eigenfunctions associated
with the continuous spectrum are not fully understood.
Recently, as a first step towards the full spectral
characterization of intermittent chaos in non-hyperbolic maps,
we have obtained
generalized eigenfunctions of the FP operator\cite{STPGDorf}
for a piecewise linear version of the Pomeau-Manneville 
map\cite{Pomeau} introduced by Artuso\cite{Artuso}. 
We continue the study of Ref.\cite{STPGDorf} for non-hyperbolic maps
with anomalous diffusion and, in this paper, we investigate 
generalized eigenvalues and the associated eigenfunctions of the FP
operator for a piecewise linear version of the
map studied by Geisel, Nierwetberg and 
Zacherl\cite{GNZ}. 
The mean square displacement of this map spreads faster than the 
normal diffusion, or it exhibits super-diffusion. 
Maps with sub-diffusion such as that studied by 
Geisel and Thomae\cite{GT} 
will be discussed elsewhere.

Note that anomalous transports such as the L\'evy 
diffusion and L\'evy flight are 
known to appear in a variety of natural systems and 
have been extensively 
studied (see \cite{AnomalousRew} and refereces therein). 
and still new maps with anomalous diffusion are 
proposed\cite{Klafter2,Aizawa}.
Then, it is worthwhile to develop a dynamical systems 
approach because
it would provide a microscopic understanding of the 
anomalous transports. This is 
one of the objectives 
of this work. 

The paper is organized as follows.  In Sec.\ref{sec.model}, we describe
a model and classes of observables and initial densities.
Evolution of statistical averages of observables is investigated in
Sec.\ref{sec.spectral} and generalized eigenfunctions of the
FP operator are derived. 
The behavior of the wavenumber-dependent eigenvalues responsible for the anomalous
diffusion is investigated in Sec.\ref{sec.Diffusion}. As one of the applications, 
`macroscopic' transports are considered  
in appropriate scaling limits. The normal diffusion is derived
if the reduced map is in the stationary regime with normal fluctuations and L\'evy 
flights are
obtained if  the reduced map is in the stationary regime with L\'evy-type fluctuations. 
And if the reduced map is in the non-stationar regime, a new transport equation is 
shown to be obtained.
Conclusions are drawn in
Sec. \ref{sec.conclusions}.


\section{Piecewise Linear Geisel-Nierwetberg-Zacherl Map}
\label{sec.model}

\begin{figure}[t]
\epsfxsize=12.0 cm
\centerline{\epsfbox{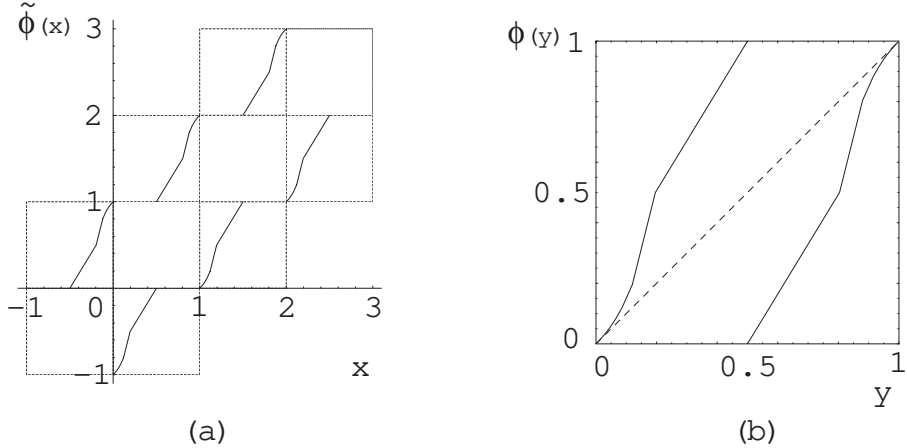}} 
\caption{(a) The piecewise linear Geisel-Nierwetberg-Zacherl map 
${\widetilde \phi}$ and (b) its reduced map $\phi$ ($\beta=2$).%
\label{fig:map}}
\end{figure}

One of the simplest maps with super-diffusion is that
introduced by Geisel, Nierwetberg and Zacherl\cite{GNZ}.
Here we consider its piecewise linear version $\tilde \phi$.
The map is defined on the whole real line (Fig.\ref{fig:map}~(a)):
${\tilde \phi}:{\bf R} \to {\bf R}$ as
\begin{equation}
{\tilde \phi}(x)= \cases{\phi(x-[x])+[x]-1 \ , &if $x-[x]\le 1/2$ \cr
\phi(x-[x])+[x]+1 \ . &if $x-[x]>1/2$ \cr
} \label{PeriodicIntMap}
\end{equation}
The symbol $[x]$ stands for the smallest integer which does not exceed $x$,
and $\phi:[0,1]\to [0,1]$ is the reduced map (cf. Fig.\ref{fig:map}~(b)) defined as follows:
\begin{equation}
\phi(y)=\cases{ \eta_k (y- \xi_k) +\xi_{k-1} \ , &$(\xi_k \le y \le
\xi_{k-1})$ \cr
\eta_k (y- 1 + \xi_k) +1-\xi_{k-1} \ , &$(1-\xi_{k-1} \le y \le
1-\xi_k )$ \cr
} \label{IntMap}
\end{equation}
where $k=1,2,\cdots$ and the variables $\xi_k$ and $\eta_k$ are
given by
\begin{equation}
\xi_{k-1}-\xi_k = {1 \over 2 \zeta\left(\beta \right)}
\left({1\over k}\right)^{\beta} \ (k=1,2,\cdots)
\ , \quad \xi_0 = {1\over 2} \ \label{IntPara1}
\end{equation}
\begin{equation}
\eta_k =\cases{\displaystyle {\xi_{k-2}-\xi_{k-1} \over
\xi_{k-1}-\xi_k} =\left({k\over k-1}\right)^{\beta}
 \ , &($k=2,3,\cdots)$ \cr
\displaystyle {1\over 1- 2 \xi_1} = \zeta\left(\beta
\right)
\ , &($k=1)$ \cr
} \label{IntPara2}
\end{equation}
In the above, $\beta >1$ is a parameter and $\zeta(s)= \sum_{n=1}^\infty
n^{-s}$ is the Riemann zeta function.
The map $\phi$ has marginal fixed points $y=0$ and $y=1$ where
the map is approximated, respectively, as 
$\phi(y) \simeq y + c_0 y^{\beta/(\beta-1)}$
and $\phi(y) \simeq y - c_0 (1-y)^{\beta/(\beta-1)}$
with a positive constant $c_0$.
When $\beta >2$, the map $\phi$ is in the stationary regime and, 
when $\beta
\le 2$, it is in the non-stationary regime.
Moreover, fluctuations in the stationary regime are normal when
$\beta >3$ and of L\'evy-type when $3 \ge \beta >2$\cite{PG-Wang}.
Those fluctuations lead to anomalous diffusion.
Note that a similar piecewise linear map was studied by Artuso\cite{Artuso}. 

The FP operator ${\hat P}$ and its adjoint ${\hat P}^*$
are defined as
\begin{eqnarray}
{\hat P} \rho(x) &=& \int_{-\infty}^{+\infty} dy \delta\left(x - {\tilde \phi}(y) \right) \rho(y)
\label{FPop} \\
{\hat P}^* A(x) &=& A\left({\tilde \phi}(x)\right) \ .
\label{KoopmanOp}
\end{eqnarray}
Thanks to the periodicity of the system, it is convenient to introduce 
the Fourier representation with respect to the integer part of the coordinate $[x]$:
\begin{eqnarray}
{\hat A}(q,y) &=& \sum_{n=-\infty}^{+\infty} e^{inq} A(n+y) \ , \qquad
{\hat \rho}(q,y) = \sum_{n=-\infty}^{+\infty} e^{inq} \rho(n+y) \ , 
\end{eqnarray}
where $y\in [ 0,1)$. 
Then, the average value of an observable $A$ at time $t$ with respect
to an initial density $\rho$ is given by 
\begin{eqnarray}
( A, {\hat P}^t \rho )
&\equiv&  ( {\hat P}^{* t} A,  \rho )
=\int_{-\pi}^{+\pi}{dq \over 2\pi} \int_0^1 dy
{\hat A}_t(q,y)^* {\hat \rho}(q,y)
\end{eqnarray}
where $(A, \rho)$ is the inner product:
\begin{equation}
(A, \rho) = \int_{-\infty}^{+\infty} dx A(x) \rho(x) 
= \int_{-\pi}^{\pi}{dq\over 2\pi}\int_0^1 dy 
{\hat A}(q,y)^* {\hat \rho}(q,y)  \ , \label{inner}
\end{equation}
and ${\hat A}_t(q,y)^*$ is the Fourier transform of ${\hat P}^{t*}A$ obeying
the recursion relation:
\begin{equation}
{\hat A}_{t+1}(q,y)^* = \cases{ e^{-iq} {\hat A}_t\left(q,\phi(y)\right)^* \ , &$0\le y <{1\over 2}$ \cr
e^{iq} {\hat A}_t\left(q,\phi(y)\right)^* \ , &${1\over 2} \le y <1$ 
}
\end{equation}
In order to extract information about generalized eigenfunctions
of the FP operator, we consider the time evolution of
the averages of observables 
with respect to a class of initial densities.

The Fourier transform of observables ${\hat A}(q,y)$ are assumed to 
be smooth at the origin and at $y=1$ in the sense
that they behave as 
${\hat A}(q,y)={\hat A}(q,0)+a_1(q) y +{\rm
O}\left(y^{\beta/(\beta-1)}\right)$ near $y=0$ and
as ${\hat A}(q,y)={\hat A}(q,1)+a_2(q) (y-1) +{\rm
O}\left((1-y)^{\beta/(\beta-1)}\right)$ 
near $y=1$, where $a_1$ 
and $a_2$  correspond, respectively, to $\partial_y{\hat A}(q,y)|_{y=0}$, 
and $\partial_y{\hat A}(q,y)|_{y=1}$.
More precisely, ${\hat A}(q,y)$ is a bounded function such that there
exist functions $a_1(q)$ and $a_2(q)$ of $q$ and inequalities
\begin{eqnarray}
&&|{\hat A}(q,y)-{\hat A}(q,0)-a_1(q) y|\le K |y|^{\beta/(\beta-1)} \\
&&|{\hat A}(q,y)-{\hat A}(q,1)-a_2(q) (y-1)|\le K' |1-y|^{\beta/(\beta-1)}  
\end{eqnarray}
hold for $0\le y \le 1$ and $-\pi \le q \le \pi$, where $K$ and $K'$
are positive constants. Let $X_O$ be a set of such
functions, then it is a Banach space with respect to the norm:
\begin{eqnarray}
\Vert A\Vert_O =&& \sum_{j=1,2}\Vert a_j\Vert_\infty +
\Vert {\hat A}\Vert_\infty
+\sup_{q, 0<y\le 1}{|{\hat A}(q,y)-{\hat A}(q,0)-a_1(q) y|\over 
|y|^{\beta/(\beta-1)}} \nonumber \\
&&+\sup_{q, 0\le y< 1}{|{\hat A}(q,y)-{\hat A}(q,1)-a_2(q) (y-1)|\over 
|1-y|^{\beta/(\beta-1)}}
\ ,
\end{eqnarray}
where $\Vert \cdot \Vert_{\infty}\equiv \sup_{q,x} |\cdot |$. The space
$X_O$ is invariant with respect to the adjoint of the FP
operator ${\hat P}^*$: namely, if $A\in X_O$, then ${\hat P}^* A\in
X_O$. Note that since $X_O$ contains a space $C^2[-\pi,\pi]\times[0,1]$ 
of twice
continuously differentiable functions and $C^2[-\pi,\pi]\times[0,1]$ 
is dense in the Hilbert space
$L^2[-\pi,\pi]\times[0,1]$ of square integrable functions, $X_O$ is 
dense in $L^2[-\pi,\pi]\times[0,1]$.

It was shown \cite{Pollicott,Ruelle} that the spectrum of the 
FP operator of expanding maps is controlled by 
the smoothness of densities. In this proof, the positivity of
the Lyapunov exponents plays an essential role. 
However, the map $\phi$ does not enjoy such a property and 
the general situation is not known. So, we restrict ourselves 
to a narrow class of piecewise constant initial densities 
in order to derive explicit expressions of
the generalized eigenfunctions of the FP operator.
More precisely, we consider a set $X_D$ of initial densities $\rho$ 
given by
\begin{equation}
{\hat \rho}(q,y)=\cases{ {\widetilde \rho}^-_k(q) \ , &$(\xi_k \le y \le
\xi_{k-1})$ \cr
{\widetilde \rho}^+_k(q) \ , &$(1-\xi_{k-1} \le y \le
1-\xi_k)$  \cr
}  \label{Initial}
\end{equation}
where $k=1,2,\cdots$ and
\begin{equation}
{\widetilde  \rho}_k^\pm(q) = \sum_{l=0}^{+\infty} \rho_l^\pm(q) 
\left({k
\over k+l}\right)^\beta \qquad {\rm with} \ \quad  
\sum_{l=0}^{+\infty} \Vert \rho_l^\pm \Vert_\infty
\theta^l <+\infty
\label{Initial2}
\end{equation}
for some constant $\theta > 1$.
The space $X_D$ is a Banach space with respect to the norm
\begin{equation}
\Vert \rho \Vert_D = \sum_{l=0}^{+\infty} \{\Vert\rho_l^+\Vert_\infty
+\Vert\rho_l^-\Vert_\infty\}
\theta^l \ ,
\end{equation}
and is invariant under the action of the FP operator
${\hat P}$: namely, if $\rho\in X_D$, then ${\hat P}\rho \in X_D$.
Note that, in contrast to the space $X_O$ of observables, the
space $X_D$ of densities is not dense in the Hilbert space 
$L^2[-\pi,\pi]\times[0,1]$ of square integrable functions.

\section{Evolution of statistical averages of observables}
\label{sec.spectral}

\subsection{Matrix elements of the resolvent}

In this section, we investigate the evolution of the
average $(A,{\hat P}^t \rho)$ of an observable $A$ with
respect to a piecewise constant initial density $\rho$.
We begin with the Neumann series of the matrix element of the
resolvent:
\begin{equation}
(A, {1\over z {\bf 1} - {\hat P}} \ \rho)= \sum_{t=0}^{+\infty}
{1\over z^{t+1}} \ (A, {\hat P}^t \rho) 
\label{Neumann}
\end{equation}
where $\bf 1$ is the identity operator. In terms of
\begin{eqnarray}
B_k^-(t,q) &=& \int_{\xi_k}^{\xi_{k-1}} dy  {\hat A}_t(q,y)^* \ , \quad {\rm and} \quad
B_k^+(t,q) = \int_{1-\xi_{k-1}}^{1-\xi_k} dy  {\hat A}_t(q,y)^* \ ,
\label{B_k}
\end{eqnarray}
each term of the series can be rewritten as
\begin{eqnarray}
(A, {\hat P}^t \rho) = \int_{-\pi}^{\pi}{dq\over 2\pi} 
\sum_{k=0}^\infty \left\{ {\widetilde \rho}^+_k(q) B^+_k(t,q)
+{\widetilde \rho}^-_k(q) B^-_k(t,q) \right\} \ , \label{NeumannEach}
\end{eqnarray}
and, hence, 
\begin{equation}
(A, {1\over z {\bf 1} - {\hat P}} \ \rho)= \int_{-\pi}^{+\pi} 
{dq\over 2\pi} 
\sum_{k=1}^{+\infty} \left\{ {\widetilde \rho}_k^+(q) {\hat B}^+_k(q,z) 
+{\widetilde \rho}_k^-(q) {\hat B}^-_k(q,z) 
\right\} \ ,
\label{MatrixElement}
\end{equation}
where ${\hat B}_k^\pm(z,q) \equiv \sum_{t=0}^\infty B_k^\pm(t,q)/z^{t+1}$.
The right-hand side of (\ref{MatrixElement}) is calculated as follows:
For example, when $k\ge 2$, eq.(\ref{IntMap}) gives 
recursion relations of $B_k^\pm(t,q)$:
\begin{eqnarray}
B_k^-(t+1,q) &=& \int_{\xi_k}^{\xi_{k-1}} dy e^{-iq}{\hat A}_t\left(q,\eta_k 
(y-\xi_k )+\xi_{k-1}\right)^* \nonumber \\
&=& {e^{-iq} \over \eta_k}
\int_{\xi_{k-1}}^{\xi_{k-2}} dy {\hat A}_t(q,y)^*
= {e^{-iq} \over \eta_k} B_{k-1}^-(t,q) , \\
B_k^+(t+1,q) &=& {e^{iq} \over \eta_k} B_{k-1}^+(t,q) \ ,
\label{B_kRecur1}
\end{eqnarray}
and, hence, we obtain
\begin{equation}
{\hat B}_k^\pm(q,z) ={1\over z} B_k^\pm(q,0) + {e^{\pm iq}\over z \eta_k } {\hat B}_{k-1}^\pm(q,z)
\label{B_kRecur2}
\end{equation}
Similarly, one has
\begin{equation}
{\hat B}_1^\pm(q,z) ={1\over z} B_1^\pm(q,0) + {e^{\pm iq} \over z \eta_1 } \sum_{k=1}^{+\infty}{\hat
B}_{k}^\mp(q,z)
\label{B_kRecur3}
\end{equation}
The recursion relations (\ref{B_kRecur2}) and (\ref{B_kRecur3})
give
\begin{eqnarray}
&&\sum_{k=1}^{+\infty} {\widetilde \rho}_k^\pm(q) {\hat B}^\pm_k(q,z)
= \Xi^\pm(q,z) +
{B_1^\pm(q,0) \Psi^\pm(q,z) \over z} \nonumber \\
&&~~~~~+ {\Psi^\pm(q,z) \over \Omega(q,z)}\left\{
{Z(ze^{\pm i q}) \over \left(z\zeta(\beta)\right)^2}\Phi^\pm(q,z) + {e^{\pm i q}\over z \zeta(\beta)}
\Phi^\mp(q,z)\right\}  
\label{BRho} 
\end{eqnarray}
where $\Xi^\pm(q,z)$, $\Phi^\pm(q,z)$, $\Psi^\pm(q,z)$, $\Omega(q,z)$ and 
$Z(z)$ are defined by:
\begin{eqnarray}
\Xi^\pm(q,z) &=&
\sum_{l=2}^{\infty} {l^\beta B_l^\pm(q,0)\over z}\left[ 
(ze^{\mp i q})^{l-1}
\Psi^\pm(q,z) - \sum_{k=1}^{l-1} {{\widetilde \rho}_k^\pm(q) 
\over k^\beta} (ze^{\mp iq})^{l-k} \right] \label{Neumann2}
\\
\Phi^\pm(q,z) &=& 
\sum_{l=1}^{\infty} \sum_{k=0}^{\infty} 
\left({e^{\pm iq} \over z}\right)^{k} \ \left({l\over k+l}\right)^\beta 
{B_l^\pm(q,0)\over z}
\label{Phi} \\
\Psi^\pm(q,z) &=& \sum_{k=1}^{\infty} {{\widetilde \rho}_k^\pm(q) 
\over k^\beta} \left({e^{\pm iq} \over z}\right)^{k-1}
\label{Psi} \\
\Omega(q,z) &=& 1 - { Z(ze^{iq}) Z(ze^{-iq}) \over \left(z \zeta(\beta)\right)^2}
\label{Omega} \\
Z(z) &=& \sum_{k=1}^{\infty} \ {z^{-k+1}\over
k^\beta}
\label{Zeta}
\end{eqnarray}
Infinite series in (\ref{Neumann2}), (\ref{Phi}), (\ref{Psi}) and 
(\ref{Zeta}) are
absolutely convergent for $|z|>1$ and, thus, are analytic there.
In addition, because of (\ref{Neumann}) and the analyticity of the 
resolvent in $|z|>1$, one finds
\begin{equation}
(A,{\hat P}^t \rho) = \oint_{|z|=r}{dz\over 2\pi i} \ z^t \
(A, {1\over z {\bf 1} - {\hat P}} \ \rho)
=\int_{-\pi}^{+\pi} 
{dq\over 2\pi} \oint_{|z|=r}{dz\over 2\pi i} \ z^t 
\ R(q,z)
\label{TimeEvol}
\end{equation}
where $R(q,z)\equiv \sum_{k=1}^\infty \{ {\widetilde \rho}_k^+(q) 
{\hat B}^+_k(q,z) + {\widetilde \rho}_k^-(q) 
{\hat B}^-_k(q,z) \}$ and
the integration path is a counter-clockwise circle 
centered at $z=0$ with radius $r(>1)$.

\subsection{Analytical properties of individual functions}

Here we derive analytical continuations of $\Phi^\pm(q,z)$,
$\Psi^\pm(q,z)$ and $Z(z)$ into the unit disk $|z|<1$.
With the aid of the formula $\Gamma(\beta)/k^\beta = \int_0^\infty
ds s^{\beta-1} e^{-ks}$, the series expression of 
$\Phi^\pm$ can be rewritten as
\begin{eqnarray}
\Phi^\pm(q,z) &=& \sum_{l=1}^{\infty}   {l^\beta B_l^\pm(q,0)\over
z \Gamma(\beta)} \int_0^\infty
ds s^{\beta-1} e^{-ls} \ \sum_{k=0}^{\infty} \left({e^{-s} \over z 
e^{\mp iq}}\right)^k  \nonumber \\
&=& \sum_{l=1}^{\infty}  {l^\beta B_l^\pm(q,0)
\over e^{\pm iq} \Gamma(\beta)}  \int_0^\infty
ds  \ {s^{\beta-1} e^{-ls} \over z e^{\mp iq}-e^{-s}}
\label{PhiAC} 
\end{eqnarray}
The above calculations are justified because the summation converges
uniformly in $s$ provided $|z|>1$.

We observe that, as a result of an inequality for bounded
observables $A$: \hfil \break
$l^\beta |B_l^\pm(q,0)|\le l^\beta (\xi_{l-1} - \xi_l)
\Vert A \Vert_\infty = \Vert A \Vert_\infty/(2 \zeta(\beta))$, 
the estimate
\begin{eqnarray}
\sum_{l=1}^{\infty}  \left| {l^\beta B_l^\pm(q,0)
\over e^{\pm iq} \Gamma(\beta)}  \int_0^\infty
ds  \ {s^{\beta-1} e^{-ls} \over z e^{\mp iq}-e^{-s}}\right|
\le 
{ \Vert A \Vert_\infty \over d(z e^{\mp iq},[0,1])}
\label{Estimate2} 
\end{eqnarray}
holds where $d(\xi,[0,1])$ is the distance between $\xi$ and the real
interval $[0,1]$.
It is obvious that each term of $\Phi^\pm$ in the right-hand side of
(\ref{PhiAC}) is analytic except a cut on an interval 
$\{z=t e^{\pm iq}: 0\le t \le 1 \}$.
Therefore, (\ref{PhiAC}) defines a function
which is analytic except the cut
and, then, is an analytical continuation of $\Phi^\pm$ from 
the outside of
the unit disk to the whole complex plane.

Similarly, the following analytical continuations are obtained:
\begin{eqnarray}
\Psi^\pm(q,z) &=&  {z e^{\mp iq} \over \Gamma(\beta)}
\sum_{l=0}^\infty \rho_l^\pm(q) 
\int_0^\infty du {u^{\beta-1} e^{-(l+1)u} \over z e^{\mp iq}-e^{-u}}
\label{PsiAC} \\
Z(z) &=& {z\over \Gamma(\beta)}
\int_0^\infty ds {s^{\beta-1} e^{-s} \over z-e^{-s}}
\ ,
\label{ZetaAC} 
\end{eqnarray}
both of which are analytic in the unit disk $|z|<1$ except the
cuts. The function $\Psi^\pm(q,z)$ has a cut on an interval 
$\{z=t e^{\pm iq}: 0\le t \le 1\}$
and $Z(z)$ has it on a real interval $[0,1]$.
In addition, the first two terms of (\ref{BRho}) have the following
analytical continuation:
\begin{eqnarray}
\Xi^\pm(q,z) + {B_1^\pm(q,0) \Psi^\pm(q,z) \over z}&=&
{\Psi^\pm(q,z) \over z}
\sum_{l=1}^{\infty} l^\beta B_l^\pm(q,0)
(z e^{\mp iq})^{l-1} \nonumber \\
&&- 
\sum_{l=2}^{\infty}\sum_{k=1}^{l-1} {l^\beta B_l^\pm(q,0) {\widetilde
\rho}^\pm_k(q) \over k^\beta e^{\pm i q(l-k)}} z^{l-k-1} 
\label{NeumannAC}
\end{eqnarray}
The series of the right-hand side are absolutely convergent
for $|z|<1$ and, thus, analytic there.
Note that (\ref{NeumannAC}) is not a simple rewrite since the series
in (\ref{Neumann2}) does not converge for $|z|<1$.

As a result, the integrand of (\ref{TimeEvol}) is meromorphic 
in the unit circle except the cuts. Its poles, if they exist, 
are given as zeros of $\Omega(q,z)$. When $q=0$, $\Omega(0,z)=0$
is equivalent to $Z(z)/z \pm \zeta(\beta)=0$. As easily seen,
$Z(z)/z - \zeta(\beta)=0$ has a unique zero at $z=1$ and an argument
similar to that given in Ref.\cite{STPGDorf} shows that
$Z(z)/z + \zeta(\beta)=0$ has a unique real zero on $(-1,0)$. 
Or $\Omega(0,z)$ has two real zeros; one is in the interval
$[-1,0]$ and the other at $1$. 
When $q\not= 0$, the behavior of $\Omega(q,z)$ is investigated
numerically with the aid of Newton's method and we found 
that $\Omega(q,z)$ may have, at most, two
simple real zeros; one in the interval $[-1,0]$ and the other 
in $[0,1]$, which we denote, respectively, as $\lambda_d^-(q)$ 
and $\lambda_d^+(q)$. Their properties will be discussed in 
Sec.\ref{sec.Diffusion} again.

\subsection{Decomposition of averages of observables}

\begin{figure}[b]
\epsfxsize=12.0 cm
\centerline{\epsfbox{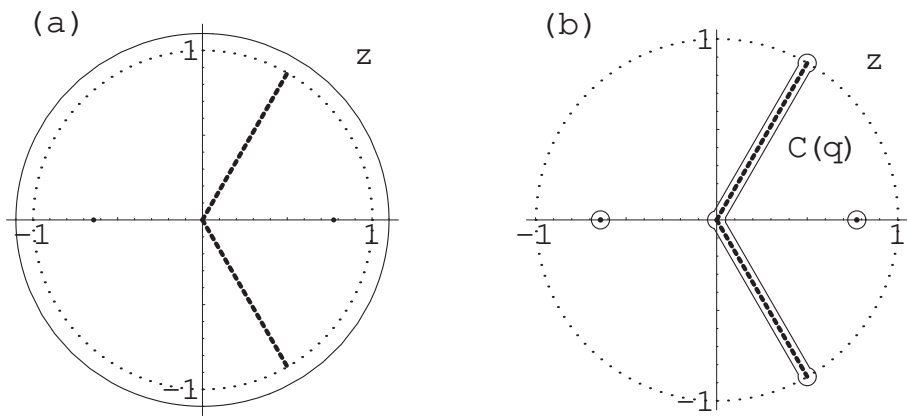}} 
\caption{Integration contours (solid curves) for individual $q$. 
Two dots and dotted lines indicate, respectively, the simple poles 
$\lambda^\pm_d(q)$ and the cuts of the integrand.  (a) The original 
circle $|z|=r (>1)$ and (b) the deformed contour consisting of 
two circles and the curve $C(q)$ surrounding two poles 
and the cuts, respectively. The directions are
counter-clockwise and the unit circle is shown as a reference.
\label{fig:contour}}
\end{figure}

Now we go back to the contour integral in (\ref{TimeEvol}). 
Note that, because of the $q$-integration, it is sufficient to 
consider $q\not =0$ terms. Then, the arguments of the previous
subsection show that, for each $q$, the integrand 
is analytic in $|z|<1$ except the simple poles at 
$z=\lambda_d^\pm(q)$ and cuts on the segments
$\{z=te^{\pm iq}:0\le~t\le~1\}$. 
Therefore, the contour integral is evaluated by deforming the 
contour from the circle $|z|=r$ to the counter-clockwise curve 
$C(q)$ shown in Fig.\ref{fig:contour} and one gets
\begin{eqnarray}
\oint_{|z|=r}{dz\over 2\pi i} \ z^t R(q,z)
= \sum_{\sigma=\pm} {\rm Res}_{\lambda_d^\sigma} 
z^t R(q,z) + \oint_{C(q)}{dz\over 2\pi i} \ z^t R(q,z)
\end{eqnarray}
where $\lambda_d^\pm(q)$ are abbreviated as $\lambda_d^\pm$ and
${\rm Res}_{\lambda_d^\pm}$ \ stands for the residue at
$z=\lambda_d^\pm$.
Firstly, we consider the contribution from the curve $C(q)$ in case
of $q>0$, which
consists of two counter-clockwise arcs 
$\gamma_{0R}=\{z:|z|=\epsilon~\&~|{\rm arg}z|<|q|\}$ and 
$\gamma_{0L}=\{z:|z|=\epsilon~\&~|\pi-{\rm arg}z|<\pi-|q|\}$,
two counter-clockwise circles $\gamma_\pm=\{z:|z-e^{\pm iq}|=\epsilon\}$,
the segment 
$C_{+l}^\epsilon=\{z=(t-i0)e^{iq}:\epsilon\le~t\le~1-\epsilon\}$ 
\break
($C_{+u}^\epsilon=\{z=(t+i0)e^{iq}:\epsilon\le~t\le~1-\epsilon\}$)
just below (above) the upper cut and
the segment $C_{-l}^\epsilon=\{z=(t-i0)e^{-iq}:\epsilon\le~t\le~1-\epsilon\}$
\break
($C_{-u}^\epsilon=\{z=(t+i0)e^{-iq}:\epsilon\le~t\le~1-\epsilon\}$) 
just below (above) the lower cut:
$$
C(q)=\gamma_{0R}\cup C_{+l}^\epsilon \cup \gamma_+ \cup C_{+u}^\epsilon 
\cup \gamma_{0L}
\cup C_{-l}^\epsilon \cup \gamma_- \cup C_{-u}^\epsilon \ .
$$
Here, $\epsilon$ stands for a small positive number, $C_{+u}^\epsilon$ 
and $C_{-u}^\epsilon$ point to the origin, and $C_{+l}^\epsilon$ 
and $C_{-l}^\epsilon$ point to the infinity.
 
Since the conditions $A\in X_O$ and $\rho\in X_D$ lead to
$$
\lim_{z\to 0}|z|^\alpha |R(q,z)|= 0 \quad (0 
< \exists \alpha <1) \ ,
$$
the contributions from $\gamma_{0R}$ and $\gamma_{0L}$ vanish 
in the limit of $\epsilon \to 0$. Similarly, the contributions
from $\gamma_{\pm}$ vanish in the same limit since
$$
\lim_{z\to e^{\pm iq}}|z-e^{\pm iq}|^{\alpha'} |R(q,z)|= 0 
\quad (\min(2-\beta,0) <\alpha' 
<1) \ .
$$
On the other hand, we have 
\begin{eqnarray}
\lim_{\epsilon\to 0}\int_{C_{+l}^\epsilon \cup C_{+u}^\epsilon}
{dz\over 2\pi i} 
\ z^t R(q,z)
= \int_{C_q^+}{dz\over 2\pi i} \ z^t \{R(q,z_l^+)-
R(q,z_u^+)\} ,
\end{eqnarray}
where $C_q^+\equiv\{z=te^{i q}:0\le t\le 1\}$ is the upper cut and 
$z_{u(l)}^+=z +(-) i e^{i q}0$. 
Eqs.(\ref{BRho}), (\ref{PhiAC}) (\ref{PsiAC}) and (\ref{ZetaAC}) 
lead to 
\begin{eqnarray}
{R(q,z_l^+)-R(q,z_u^+) \over 2\pi i} = &&N_+(q,z)\left[\Delta\Psi^+(q,z)
+{\Psi^-(q,z)\over ze^{iq}\zeta(\beta)}\right] \nonumber \\
&&\mskip 50 mu \times ze^{-iq} \left[\Delta\Phi^+(q,z)
+{\Phi^-(q,z)\over ze^{-iq}\zeta(\beta)}\right]
\end{eqnarray}
where 
\begin{eqnarray}
N_+(q,z)&=& { \left[-\ln(ze^{-iq})\right]^{\beta-1} \over
\Gamma(\beta) \Omega(q,z_l^+) \Omega(q,z_u^+)} \\
\Delta\Psi^+(q,z) &=& {\Gamma(\beta) \left[\Omega(q,z_u^+) 
\Psi^+(q,z_l^+) - \Omega(q,z_l^+) 
\Psi^+(q,z_u^+) \right] \over 2\pi i ze^{-iq}
\left[-\ln(ze^{-iq})\right]^{\beta-1}} \\
\Delta\Phi^+(q,z) &=& {\Gamma(\beta) \left[\Omega(q,z_u^+) 
\Phi^+(q,z_l^+) - \Omega(q,z_l^+) 
\Phi^+(q,z_u^+) \right] \over 2\pi i ze^{-iq}
\left[-\ln(ze^{-iq})\right]^{\beta-1}}  \ .
\end{eqnarray}
The contributions from the cut $C_q^+$ with $q<0$ as well as 
from the other cut $C_q^- \equiv \{z=te^{-iq}:
0\le t \le 1\}$ are evaluated in a similar way and finally one
obtains
\begin{eqnarray}
\lim_{\epsilon\to 0}\int_{C_{\pm l}^\epsilon \cup C_{\pm u}^\epsilon}
{dz\over 2\pi i} 
\ z^t R(q,z)
=\int_{C_q^\pm} dz \ \Bigl(A,F_\pm(q,z)\Bigr) \ z^t  \
\Bigl({\widetilde F}_\pm(q,z),\rho \Bigr)
\end{eqnarray}
The functional $\Bigl(A,F_\pm(q,z)\Bigr)$ is defined by
\begin{eqnarray}
&&\Bigl(A,F_\pm(q,z)\Bigr) = \int_0^1 dy {\hat A}(q,y)^* g_{qz}^\pm(y)
\ , \\
&&\ \ g_{qz}^\pm(y) = \sum_{l=1}^\infty { e^{\pm iq} l^\beta \chi_l^\mp(y) \over 
\zeta(\beta) \Gamma(\beta)} \int_0^\infty ds {s^{\beta-1}
e^{-ls}\over ze^{\pm iq}-e^{-s}} + {\Omega_{qz}^\pm \over z} 
\sum_{l=1}^\infty l^\beta \chi_l^\pm(y) (ze^{\mp iq})^l 
\nonumber \\
&&\ \ \mskip 75mu +\sum_{l=1}^\infty l^\beta \chi_l^\pm(y)
{e^{\mp i q} Z(ze^{\pm iq}) \over ze^{\pm iq} 
\zeta(\beta)^2 \Gamma(\beta)} \int_0^\infty ds \ {\mathcal P}{s^{\beta-1}
e^{-ls}\over ze^{\mp iq}-e^{-s}} \ , \\
&&\ \ \chi_l^-(y) = \chi_l^+(1-y) = \cases{ 1 \  
&\ ($\xi_l < y \le \xi_{l-1}$) \cr
0 \  &\ (otherwise)} \ , \\
&&\ \ \ \ \ \Omega_{qz}^\pm = 1 - {Z(ze^{\pm iq}) \over ze^{\pm iq}\zeta(\beta)^2 
\Gamma(\beta)} \int_0^\infty ds \ {\mathcal P}{s^{\beta-1}
e^{-s}\over ze^{\mp iq}-e^{-s}} \ ,
\end{eqnarray}
where ${\mathcal P}$ stands for Cauchy's principal value. Note that
$ze^{\mp iq}$ is real if $z\in C_q^\pm$.
Similarly, $\Bigl({\widetilde F}_\pm(q,z),\rho \Bigr)$ is given by
\begin{eqnarray}
&&\Bigl({\widetilde F}_\pm(q,z),\rho \Bigr)= n_{qz}^\pm
\biggl\{ \sum_{j=0}^\infty {\rho_j^{\pm}(q) Z(ze^{\pm iq})\over 
ze^{\pm iq} \zeta(\beta)^2
\Gamma(\beta)}\int_0^\infty ds \ {\mathcal P}{s^{\beta-1} 
e^{-(j+1)s}\over 
ze^{\mp iq}-e^{-s}} \nonumber \\
&&\mskip 75 mu +\Omega_{qz}^\pm \sum_{j=0}^\infty \rho_j^{\pm}(q) 
(ze^{\mp iq})^j
+\sum_{j=0}^\infty {\rho_j^{\mp}(q)\over 
\zeta(\beta)\Gamma(\beta)}
\int_0^\infty ds {s^{\beta-1} e^{-(j+1)s}\over ze^{\pm iq}-e^{-s}}
\biggr\} \ , \\
&&\mskip 10mu \ n_{qz}^\pm = {[-\log ze^{\mp iq}]^{\beta-1} \over \Gamma(\beta)
\Omega(q,z+ie^{\pm iq}0) \Omega(q,z-ie^{\pm iq}0)} \ .
\end{eqnarray}

The pole contributions are calculated as above and one has
\begin{eqnarray}
\sum_{\sigma=\pm} {\rm Res}_{\lambda_d^\sigma} 
z^t R(q,z) &=& 
\sum_{\sigma=\pm} {\lambda_d^\sigma(q)}^t \lim_{z\to \lambda_d^\sigma(q)}
\left(z-\lambda_d^\sigma(q)\right) R(q,z) \nonumber \\
&=& \sum_{\sigma=\pm} \Bigl(A,F_d^\sigma(q)\Bigr) \ {\lambda_d^\sigma(q)}^t  
\ \Bigl({\widetilde F}_d^\sigma(q),\rho \Bigr) \ ,
\end{eqnarray}
where the functionals are given by
\begin{eqnarray}
\Bigl(A,F_d^\sigma(q)\Bigr) &=& \int_0^1 dy {\hat A}(q,y)^* g_d^\pm(q,y)
\ , \\
g_d^\pm(q,y) &=& \sum_{l=1}^\infty {Z(\lambda_d^\pm e^{iq}) l^\beta 
\chi^+_l(y)\over \lambda_d^\pm e^{iq}\zeta(\beta)\Gamma(\beta)}
\int_0^\infty ds {s^{\beta-1} e^{-l s}\over \lambda_d^\pm e^{-iq}
-e^{-s}} \nonumber \\
&&~~+ \sum_{l=1}^\infty {e^{2iq} l^\beta 
\chi^-_l(y)\over \Gamma(\beta)}
\int_0^\infty ds {s^{\beta-1} e^{-l s}\over \lambda_d^\pm e^{iq}
-e^{-s}} \ , \\
\Bigl({\widetilde F}_d^\sigma(q),\rho \Bigr) &=&
\sum_{j=0}^\infty {e^{-iq} \rho_j^+(q)\over \Omega'(q,\lambda_d^\pm)
\zeta(\beta)\Gamma(\beta)} \int_0^\infty ds {s^{\beta-1} e^{-(j+1)s}
\over \lambda_d^\pm e^{-iq} -e^{-s}} \nonumber \\
&&+\sum_{j=0}^\infty {\lambda_d^\pm \rho_j^-(q)\over \Omega'(q,\lambda_d^\pm)
Z(\lambda_d^\pm e^{iq})\Gamma(\beta)} \int_0^\infty ds {s^{\beta-1} e^{-(j+1)s}
\over \lambda_d^\pm e^{iq} -e^{-s}} \ .
\end{eqnarray}
In the above, $\lambda_d^\pm(q)$ is abbreviated as $\lambda_d^\pm$ and 
$\Omega'(q,z)$ stands for the derivative of $\Omega(q,z)$ with 
respect to $z$.

In summary, we have derived
\begin{eqnarray}
(A, {\hat P}^t \rho) &=& \int_{-\pi}^\pi {dq \over 2\pi}
\sum_{\sigma=\pm} \Bigl(A,F_d^\sigma(q)\Bigr) \ {\lambda_d^\sigma(q)}^t  
\ \Bigl({\widetilde F}_d^\sigma(q),\rho \Bigr) \nonumber \\
&&+ \int_{-\pi}^\pi {dq \over 2\pi} \ \sum_{\alpha=\pm} \ 
\int_{C_q^\alpha} dz \ (A,F_\alpha(q,z)) \ z^t  
\ ({\widetilde F}_\alpha(q,z),\rho )
\label{TimeEvol5}
\end{eqnarray}

\subsection{Generalized eigenfunctions of the Frobenius-Perron
operator}

We remark that, for any bounded linear functional $F$ over
$X_O$, $({\hat P}^*A,F)$ ($A\in X_O$) is well defined due to the
invariance of $X_O$ under ${\hat P}^*$ and that the operator ${\hat
P}$ can be extended via $(A,{\hat P}F)\equiv ({\hat P}^*A,F)$ to the
space of all bounded linear functionals over $X_O$ (the dual space
of $X_O$).  
The linear
functionals $F_d^\pm$ and $F_\pm(q,z)$ are
bounded and satisfy the following relations for any $A\in X_O$:
\begin{eqnarray}
(A,{\hat P} F_d^\pm(q)) &\equiv& ({\hat P}^* A,F_d^\pm(q)) = 
\lambda_d^\pm(q)(A,F_d^\pm(q)) \ , \\
(A,{\hat P} F_\pm(q,z)) &\equiv& ({\hat P}^* A,F_\pm(q,z)) = 
z (A,F_\pm(q,z))
\ . \ \ \ (z\in C_q^\pm)
\label{REigenRel}
\end{eqnarray}
Namely, they are 
eigenfunctions of ${\hat P}$ in
the generalized sense\cite{GeneralSpectrum}.

Similarly, the operator ${\hat P}^*$ can be extended to the dual space
of $X_D$ and the functionals ${\tilde F}_d^\pm$, ${\tilde F}_l(q,z)$,
${\tilde F}_u(q,z)$ are its generalized eigenfunctions:
\begin{eqnarray}
({\hat P}^*{\tilde F}_d^\pm(q),\rho)&=&({\tilde F}_d^\pm(q),{\hat
P}\rho)= \lambda_d^\pm(q)({\tilde F}_d^\pm(q),\rho) \ , \\ 
({\hat P}^*{\tilde F}_\pm(q,z),\rho)&=&({\tilde F}_\pm(q,z),
{\hat P}\rho)
=z ({\tilde F}_\pm(q,z),\rho) \ , \ \ \ (z\in C_q^\pm)
\label{LEigenRel}
\end{eqnarray}
Sometimes, eigenfunctions of ${\hat P}$ and ${\hat P}^*$ are 
referred to, respectively, as right and left eigenfunctions of ${\hat
P}$.

It is then interesting to reinterpret (\ref{TimeEvol5}) from this
point of view. One has
\begin{eqnarray}
(A, \rho) &=& \int_{-\pi}^\pi {dq \over 2\pi}
\sum_{\sigma=\pm} \Bigl(A,F_d^\sigma(q)\Bigr)   
 \Bigl({\widetilde F}_d^\sigma(q),\rho \Bigr) \nonumber \\
&&+ \int_{-\pi}^\pi {dq \over 2\pi} \ \sum_{\alpha=\pm} \ 
\int_{C_q^\alpha} dz \ (A,F_\alpha(q,z)) 
({\widetilde F}_\alpha(q,z),\rho )
\label{Complete} \\
(A, {\hat P} \rho) &=& \int_{-\pi}^\pi {dq \over 2\pi}
\sum_{\sigma=\pm} \Bigl(A,F_d^\sigma(q)\Bigr) \ {\lambda_d^\sigma(q)}  
\ \Bigl({\widetilde F}_d^\sigma(q),\rho \Bigr) \nonumber \\
&&+ \int_{-\pi}^\pi {dq \over 2\pi} \ \sum_{\alpha=\pm} \ 
\int_{C_q^\alpha} dz \ (A,F_\alpha(q,z)) \ z
\ ({\widetilde F}_\alpha(q,z),\rho )
\label{SpectralDecom}
\end{eqnarray}
The first relation (\ref{Complete}) corresponds to the completeness
of the set of left eigenfunctions $\{ F_d^\pm(q), F_\pm(q,z) \}$ and
that of right ones $\{ {\tilde F}_d^\pm(q), {\tilde F}_\pm(q,z) \}$. 
The second relation (\ref{SpectralDecom}) 
can be considered as a spectral decomposition of ${\hat P}$ in terms of
those eigenfunctions, which correspond, for each $q$, to two isolated 
eigenvalues $\lambda_d^\pm(q)$ and continuous spectra on $C_q^\pm$.
These relations are regarded as an extension to the
FP operator of the generalized spectral decomposition in
the sense of Ref.\cite{GeneralSpectrum}, which was originally
formulated for self-adjoint and unitary operators.  Such extended
spectral decompositions are widely used to study statistical
properties of dynamical systems (see \cite{PGtext,DetDif3}
and references therein).

\section{Transports}
\label{sec.Diffusion}

For systems exhibiting deterministic diffusion with
diffusion coefficient $D$\cite{PGtext,Dorftext,DetDif2,DetDif3}, 
the FP operator has an eigenvalue $z=z_q$
such that $z_q\simeq 1-Dq^2$ for small wavenumber $|q|$ and 
it fully characterizes the diffusion in a scaling limit. 
A similar behavior is expected for the map ${\widetilde
\phi}$. We then investigate an eigenvalue $\lambda_d^+(q)$ 
for $|q|\ll 1$ and appropriate scaling limits.

\subsection{Case of $\beta >3$}

\begin{figure}[t]
\epsfxsize=12.0 cm
\centerline{\epsfbox{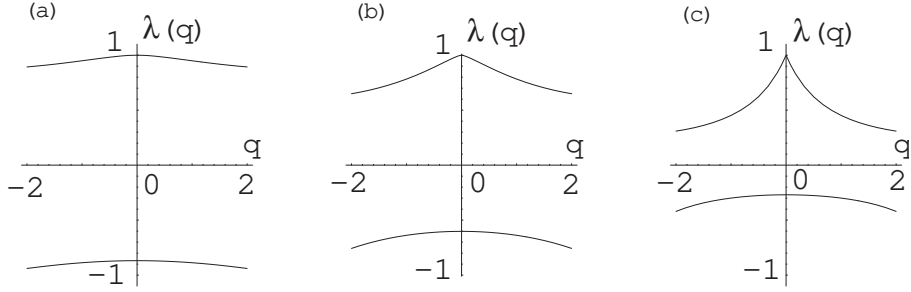}} 
\caption{The $q$-dependence of the eigenvalues of the FP
operator. The upper branch is $\lambda_d^+(q)$ and the lower branch is
$\lambda_d^-(q)$. (a) $\beta=4$, (b) $\beta=2.5$ and (c) 
$\beta=1.7$.%
\label{fig.dispersion}}
\end{figure}

When $|q|\ll 1$, $|1-\lambda_d^+ e^{\pm iq}|\ll 1$ and 
it is sufficient to retain the largest terms of $\Omega(q,z)$ 
with respect to $|1-z e^{\pm iq}|$. For $\beta >3$, they are
found to be 
\begin{eqnarray}
\Omega(q,z)\simeq&& 
{2 \zeta(\beta-1)\over \zeta(\beta)}(z\cos q -1)
-
\left({\zeta(\beta-1)\over \zeta(\beta)}\right)^2
(1+z^2-2z\cos q) \nonumber \\
&&-{\zeta(\beta-1)+\zeta(\beta-2)\over \zeta(\beta)}
(1-2z \cos q + z^2 \cos 2q) \ .
\end{eqnarray}
Then, the equation $\Omega(q,\lambda_d^+)=0$ has a solution
\begin{eqnarray}
\lambda_d^+ \simeq 1-{1\over 2}\left[{\zeta(\beta-2)\over \zeta(\beta-1)
}-{\zeta(\beta-1)\over \zeta(\beta)}\right] \ q^2 \ ,
\end{eqnarray}
which implies that the map exhibits ordinary diffusion.
In Fig.\ref{fig.dispersion}~(a), the $q$-dependence of 
numerically calculated $\lambda_d^\pm(q)$ for $\beta=4$ is shown.
The $q^2$-behavior of $\lambda_d^+(q)$ near $q=0$ is clearly seen.

The appearance of ordinary diffusion can be seen more
explicitly by considering a scaling limit, where one sets 
$n=L_\xi X$, $t=L_\tau T$ and the limit of $L_\xi \to +\infty$ is
taken while assuming a functional relation between $L_\xi$ and
$L_\tau$.  Accordingly, the observable $A(L_\xi X)$ is
assumed to be a well-behaved function of $X$. Then, one has
\begin{equation}
{1\over L_\xi}{\hat A}(Q/L_\xi)\equiv {1\over L_\xi} 
\sum_{n=-\infty}^\infty
e^{inQ/L_\xi} A(n) 
\simeq {\hat A}^{\rm (cg)}(Q) 
\end{equation}
where ${\hat A}^{\rm (cg)}(Q) \equiv \int_{-\infty}^\infty dX
e^{i QX} A(L_\xi X)$ is the Fourier transform of the 
coarse-grained observable. Similarly, the Fourier transform of the 
probability distribution is assumed to be constant on each unit interval and
is a slowly varying function of the site coordinate $[x]$. Then, in terms of the 
Fourier transform of the coarse-grained density ${\hat \rho}^{\rm (cg)}$,
one has $\rho_l^\pm(Q/L_\xi)=\delta_{l,0} {\hat \rho}^{\rm (cg)}(Q)$. Hence, 
\begin{eqnarray}
&&(A,{\hat P}^{L_\tau T} \rho) \simeq \int_{-\infty}^\infty {dQ\over 2\pi}
{\hat A}^{\rm (cg)}(Q)^* {\hat \rho}^{\rm (cg)}(Q) \Biggl\{
\sum_{\sigma=\pm} \Bigl(e,F_d^\sigma(q)\Bigr) \ {\lambda_d^\sigma(q)}^{L_\tau T}   
\ \Bigl({\widetilde F}_d^\sigma(q),e \Bigr) \nonumber \\
&& \ + \ \sum_{\alpha=\pm} \ e^{\alpha iq (L_\tau T+1)} \
\int_0^1 ds \ (e,F_\alpha(q,s e^{\alpha iq})) 
\ s^{L_\tau T} 
\ ({\widetilde F}_\alpha(q,s e^{\alpha iq}),e ) 
\Biggr\}_{q={Q\over L_\xi}}
\end{eqnarray}
where $e$ is a function of $n$ and $y$ with the Fourier transform ${\hat e}(q,y)
= 1$. In order to have a non-trivial scaling limit, one must
have $L_\tau \propto L_\xi^2$ and here we take $L_\tau = L_\xi^2$.
Then, one has
\begin{equation}
\lim_{L_\xi\to \infty \atop L_\tau = L_\xi^2}
\left(e,F_d^+\left({\scriptstyle {Q\over L_\xi}}\right)\right) 
\ {\lambda_d^+
\left({\scriptstyle {Q\over L_\xi}}\right)}^{L_\tau T}   
\ \left({\widetilde F}_d^+ 
\left({\scriptstyle {Q\over L_\xi}}\right),e \right)
= e^{-D^N_\beta Q^2 T} 
\end{equation}
where $D^N_\beta = {1\over 2}\left[{\zeta(\beta-2)\over \zeta(\beta-1)
}-{\zeta(\beta-1)\over \zeta(\beta)}\right]$ is the diffusion coefficient.
In this scaling limit,
the contribution from $\lambda_d^-$ vanishes as 
$\sup_q|\lambda_d^-(q)|<1$.
This is also the case for the continuous
spectrum as $k(s,Q)\equiv {\displaystyle \lim_{L_\xi \to \infty}}
(e,F_\alpha(q,s e^{\alpha iq})) 
({\widetilde F}_\alpha(q,s e^{\alpha iq}),e )|_{q={Q\over L_\xi}}$
is a bounded function of $s$:
\begin{eqnarray}
&&\lim_{L_\xi\to \infty \atop L_\tau = L_\xi^2}\left|
\int_0^1 ds \ \left(e,F_\alpha\left({\scriptstyle {Q\over L_\xi}},
s e^{\alpha iq}\right)\right) 
\ s^{L_\tau T} 
\ \left({\widetilde F}_\alpha\left({\scriptstyle {Q\over L_\xi}},
s e^{\alpha iq}\right),e \right)\right| \nonumber \\
&&\mskip 40 mu \le \sup_{s,Q}|k(s,Q)|  \lim_{L_\tau \to \infty}
\int_0^1 ds s^{L_\tau T} =0 \ .
\end{eqnarray}
In short, the scaling limit of the average value
$\langle A \rangle_T \equiv \lim_{L_\xi\to \infty \atop L_\tau = L_\xi^2}
\langle A , {\hat P}^{L_\tau T} \rho \rangle$ of $A$ at time $T$ is 
given by
\begin{equation}
\langle A \rangle_T = \int_{-\infty}^\infty {dQ\over 2\pi} 
{\hat A}^{\rm (cg)}(Q)^* \ e^{-D^N_\beta Q^2 T} \ 
{\hat \rho}^{\rm (cg)}(Q)
\ ,
\end{equation}
which is nothing but the solution of the diffusion equation.

\subsection{Case of $3 \ge \beta >2$}

When $\beta=3$, the largest terms of $\Omega(q,z)$ 
with respect to $|1-z e^{\pm iq}|$ are
\begin{eqnarray}
\Omega(q,z)\simeq&& 
{2 \zeta(2)\over \zeta(3)}(z\cos q -1)
+\sum_{\sigma=\pm}{(1-ze^{\sigma iq})^2 \log(1-ze^{\sigma iq}) \over 2 
\zeta(3)} \ ,
\end{eqnarray}
from which one obtains the following solution of 
$\Omega(q,\lambda_d^+)=0$:
\begin{equation}
\lambda_d^+ \simeq 1 - {q^2 |\log q^2| \over 4 \zeta(2)} \ .
\end{equation}
Similarly, when $2<\beta <3$, the leading term of $\Omega(q,z)$ 
is
\begin{eqnarray}
\Omega(q,z)\simeq&& 
{2 \zeta(\beta-1)\over \zeta(\beta)}(z\cos q -1)+\sum_{\sigma=\pm}
{\pi e^{-i\pi \beta} (1-ze^{\sigma iq})^{\beta-1} 
\over \Gamma(\beta)\zeta(\beta)\sin \pi\beta}
 \ ,
\end{eqnarray}
and the approximate solution of $\Omega(q,\lambda_d^+)=0$ is
\begin{equation}
\lambda_d^+ \simeq 1 - {\pi |q|^{\beta-1} \over 2 \zeta(\beta-1)
\Gamma(\beta)\left|\cos {\pi \beta \over 2}\right|} \ .
\end{equation}
In both cases, the system exhibits anomalous diffusion.
The $q$-dependence of numerically calculated $\lambda_d^\pm$
for $\beta=2.5$ is shown in Fig.\ref{fig.dispersion}~(b) and 
a $q$-dependence different from that for the normal diffusion is 
seen in the figure.

Now we consider a scaling limit for $2<\beta <3$, where we set 
$n=L_\xi X$, $t=L_\tau T$ and take the limit of $L_\xi \to +\infty$
while keeping $L_\tau=L_\xi^{\beta-1}$. Then, the contributions 
from the eigenvalue $\lambda_d^-$ and continuous spectrum vanish 
because of the same reason as in the previous subsection, and
in terms of 
the coarse-grained observable ${\hat A}^{\rm (cg)}(Q)$ and probability 
density ${\hat \rho}^{\rm (cg)}(Q)$ introduced
previously, one has
\begin{equation}
\langle A \rangle_T \equiv \lim_{L_\xi\to \infty \atop L_\tau = 
L_\xi^{\beta-1}}
\langle A , {\hat P}^{L_\tau T} \rho \rangle= 
\int_{-\infty}^\infty {dQ\over 2\pi} 
{\hat A}^{\rm (cg)}(Q)^* \ e^{-D^L_\beta |Q|^{\beta-1} T} \ 
{\hat \rho}^{\rm (cg)}(Q)
\ ,
\end{equation}
where $D^L_\beta ={\pi \over 2 \zeta(\beta-1)
\Gamma(\beta)\left|\cos {\pi \beta \over 2}\right|}$.

\subsection{Case of $2 \ge \beta > 1$}

When $\beta =2$, the leading term of $\Omega(q,z)$ 
with respect to $|1-z e^{\pm iq}|$ is 
\begin{eqnarray}
\Omega(q,z)\simeq 
{1-ze^{iq}\over \zeta(2)}\log{1-ze^{iq}\over -ze^{iq+1}}
+ {1-ze^{-iq}\over \zeta(2)}\log{1-ze^{-iq}\over -ze^{-iq+1}}
 \ .
\end{eqnarray}
By setting $-ze^{-iq} = e^{i\pi}+r e^{i \theta}$ and 
$-ze^{iq} = e^{i\pi}+r e^{i(2\pi - \theta)}$ ($0<\theta <2\pi$),
one finds the approximate solution of $\Omega(q,\lambda_d^+)=0$:
\begin{equation}
\lambda_d^+ \simeq 1- {\pi |q| \over 2 \left|\log|q| \right|}
 \ . \qquad (|q| \ll 1)
\end{equation}
On the other hand, when $1<\beta <2$, the leading term of
$\Omega(q,z)$ is
\begin{eqnarray}
\Omega(q,z)\simeq {-\pi e^{-i\pi \beta} \over \zeta(\beta) \Gamma(\beta)
|\sin \pi \beta|}\left\{(1-ze^{iq})^{\beta-1}+ 
(1-ze^{-iq})^{\beta-1}\right\} \ ,
\end{eqnarray}
and, by the same way as above, 
one finds that $\Omega(q,\lambda_d^+)=0$ admits a solution
\begin{equation}
\lambda_d^+ = {\sin \left({\pi\over 2(\beta-1)}\right) \over 
\sin \left({\pi \over 2(\beta-1)}-|q|\right)}
\simeq 1- \left| \cot{\pi \over 2(\beta-1)}\right| \ |q|
 \ , \qquad (|q| \ll 1)
\end{equation}
when $3/2 <\beta <2$ and has no solution when $\beta\le 3/2$.
The $q$-dependence of numerically calculated $\lambda_d^\pm$
for $\beta=1.7$ is shown in Fig.\ref{fig.dispersion}~(c), where
$|q|$-linear behavior of $\lambda_d^+$ is seen near $q=0$.
Moreover, for $\beta=1.1$, Newton's method fails to find a solution
of $\Omega(q,z)=0$ and this is consistent with the absence of 
the solution.

Finally, we consider a scaling limit for $1<\beta <2$, where we set 
$n=L_\xi X$, $t=L_\tau T$ and take the limit of $L_\xi \to +\infty$
while keeping $L_\tau=L_\xi$. As before, the contribution 
from the eigenvalue $\lambda_d^-$ vanishes. 
In this case, however, the continuous spectra do contribute and, after
a tedious but straightforward calculation, one obtains
\begin{eqnarray}
\langle A \rangle_T \equiv \lim_{L_\xi\to \infty \atop L_\tau = 
L_\xi}
\langle A , {\hat P}^{L_\tau T} \rho \rangle 
&=& 
\int_{-\infty}^\infty {dQ\over 2\pi} 
{\hat A}^{\rm (cg)}(Q)^* \ {\hat \rho}^{\rm (cg)}(Q)
\ \biggl\{ \cos QT \nonumber \\
&& + \int_\Gamma {dz\over 2\pi}
{e^{-|Q|T z}\over z^2+1} {(z-i)^{\beta-1}-(z+i)^{\beta-1}\over 
(z-i)^{\beta-1}+(z+i)^{\beta-1}}\biggr\}\ 
\ , \label{NewTran}
\end{eqnarray}
where $\Gamma=\{ z=t+i(1+\epsilon)| t:+\infty \to 0\}\cup 
\{z=(1+\epsilon)e^{i\theta}|\theta:{\pi \over 2}
\to {3\pi \over 2}\} \cup$ $\{ z=t-i(1+\epsilon)|  t:0\to +\infty \}$ 
is the integration contour 
with $\epsilon$ an arbitrary positive constant, and the branch cut 
of $(z\mp i)^{\beta-1}$ is $\{ z=t\pm i | t:0\to +\infty\}$. 

\section{CONCLUSIONS}
\label{sec.conclusions}

We have studied, for a piecewise linear approximation
of Geisel-Nierwetberg-Zacherl map, the evolution of statistical 
averages of bounded observables which are differentiable at the 
origin with
respect to piecewise constant initial densities which have certain 
smoothness at the origin.
Because of the periodicity of the system, the average value is 
expressed by a superposition of Fourier components and the time
evolution of each component is governed by two simple eigenvalues
$\lambda_d^\pm(q)$ and continuous spectra $C_q^\pm =\{z=te^{\pm iq}:
0\le t \le 1\}$ of the FP operator. 
The corresponding generalized eigenfunctions are explicitly
constructed. 

The main difference from the hyperbolic systems with 
deterministic diffusion is the existence of the continuous
spectra starting from the unit circle for each wavenumber $q$.
In spite of this fact, when $\beta > 2$, the transport in an 
appropriate scaling limit is entirely controlled by a simple 
eigenvalue $\lambda_d^+(q)$ located near 1 and the average value
of the coarse-grained description $\langle A \rangle_T$ obeys
\begin{equation}
\langle A \rangle_T = 
\int_{-\infty}^\infty {dQ\over 2\pi} 
{\hat A}^{\rm (cg)}(Q)^* \ e^{-D_\beta |Q|^\mu T} \ 
{\hat \rho}^{\rm (cg)}(Q)
\ , \label{Levy2}
\end{equation}
where $\mu=2$ and $D_\beta=D^N_\beta$ for $\beta>3$, and
$\mu=\beta-1$ and $D_\beta=D^L_\beta$ for $3>\beta>2$.
In other words, when the reduced map is in the stationary regime, transports 
in the scaling limits are governed by the eigenvalues.  And the normal diffusion 
appears if the fluctuations are normal and the L\'evy flight appears if the fluctuations
are of L\'evy-type. 

Here a remark is in order. As is well known, in periodic maps with marginal fixed
points of Geisel-Nierwetberg-Zacherl type, 
the mean square displacement is well-defined
and grows faster than the normal diffusion. Indeed, with the aid of (\ref{TimeEvol5}), 
one recovers a known result on the mean square displacement:
\begin{equation}
\langle ([x]-\langle [x]\rangle_t )^2 \rangle_t \propto \cases{
t &$\beta>3$ \cr
t \log t &$\beta=3$ \cr
t^{4-\beta} &$3>\beta>2$ \cr
t^2 &$2>\beta>1$
} \ ,
\end{equation}
where $\langle \cdots \rangle_t$ is the average at time $t$ with 
respect to an initial distribution $\rho(x)=\delta_{[x], n_0}$ for some
$n_0$. On the other hand, the mean square displacement is ill-defined 
for L\'evy flights\cite{LevyFlight}. However, there is no contradiction between the two 
conclusions because the ill-definedness of the mean square displacement 
comes from the scaling limit.  

The situation changed for $2> \beta >1$ or when the reduced map is in the
non-stationary regime. Then the `macroscopic' transport is described by 
(\ref{NewTran}), the first term of which represents ballistic transport, while the second
term corresponds to a more complicated process.

Note that the space $X_D$ of initial densities is not dense
in $L^2[0,1]$ and we will discuss elsewhere a complementary 
description in terms of initial densities which form a dense
subset of $L^2[0,1]$.

\section*{Acknowledgments}

{\baselineskip=19pt

The authors are grateful to Professors Y. Aizawa, J. Klafter and R. Klages
for fruitful discussions and comments. 
P.G. is financially supported by the National
Fund for Scientific Research (FNRS Belgium).
This  work is supported by a Grant-in-Aid for
Scientific Research (C) from JSPS
and by a Grant-in-Aid for Scientific Research of Priority Areas 
``Control of Molecules in Intense Laser Fields'' from 
the Ministry of Education, Culture, Sports, Science and 
Technology of Japan.

}

\end{document}